# On the causes of subject-specific citation rates in Web of Science.

Werner Marx [1] und Lutz Bornmann [2]

[1] Max Planck Institute for Solid State Research, Heisenbergstraβe 1, D-70569 Stuttgart, Germany. Email: w.marx@fkf.mpg.de

[2] Administrative Headquarters of the Max Planck Society, Division for Science and Innovation Studies, Hofgartenstr. 8, 80539 Munich, Germany. Email: bornmann@gv.mpg.de

Abstract: It is well known in bibliometrics that the average number of citations per paper differs greatly between the various disciplines. The differing citation culture (in particular the different average number of references per paper and thereby the different probability of being cited) is widely seen as the cause of this variation. Based on all Web of Science (WoS) records published in 1990, 1995, 2000, 2005, and 2010 we demonstrate that almost all disciplines show similar numbers of references in the appendices of their papers. Our results suggest that the average citation rate is far more influenced by the extent to which the papers (cited as references) are included in WoS as linked database records. For example, the comparatively low citation rates in the humanities are not at all the result of a lower average number of references per paper but are caused by the low fraction of linked references which refer to papers published in the core journals covered by WoS.

The number of citations is often taken as a measure of attention a paper, a journal, a researcher or an institute has attracted. Although citation counts are no ultimate scale of the "true" quality of publications, they can be taken as a kind of proxy data for quantifying research performance. Citation-based impact indicators reflect strengths and shortcomings and are therefore frequently used for research evaluation purposes. However, the number of citations of a paper is not very meaningful on its own.

As the five graphs in figure 1 based on the Web of Science (WoS) show for the publication years 1990, 1995, 2000, 2005, and 2010 the average numbers of citations per paper (the grey bars in the graphs) differ greatly between the major subject areas (e.g. between the natural sciences and the humanities). Whereas, for example, in the year 1990 a paper in the humanities was cited 1.7 times on average, the average citation rate in the natural sciences is 24.7 citations per paper. If one compares the average citation rates of the various natural science disciplines, one again finds a bandwidth of a factor of about ten (http://thomsonreuters.com/essential-science-indicators/). The citation rates vary just as markedly between the sub-disciplines within the natural science disciplines (Neuhaus et al., 2009).

The cause of the great variation in the average number of citations per paper is often given as the differing citation culture of the various disciplines. Garfield (1979) suggests that the most accurate measure of citation potential is the average number of references per paper within a given field. This view attributes the discipline-specific citation rates to the different probability that a particular paper will be cited as a direct consequence of the different average number of references per paper. In this connection one should, however, differentiate between references comprising papers which have appeared in WoS source journals (the core journals covered by WoS) and references involving papers which have been published in other journals or comprising books, book chapters, Internet links or patents. The former references correspond to the so-called active or linked references. Like the citing papers, these papers are stored as database documents (database records). In other words: Only a particular fraction of the cited papers (the references) is identical with the citing papers (the WoS database records) and linked with these.

We would like to demonstrate that the amount of references linked with the corresponding database records and searchable as such plays a greater role than the specific citation culture. This factor has already been mentioned in earlier papers dealing with the subject-dependency of journal-related bibliometric indicators (Althouse et al., 2009; Lancho-Barrantes et al., 2010). Moed (2005, pp.120-131) analysed the WoS coverage as a function of the research areas based on the source items processed in the year 2002. He determined the amount of linked references to estimate the subject-specific differences in journal coverage and in the overall coverage of the relevant literature (i.e., books and other non-journal literature included).

For the present study the relevant data was collected and analysed – in contrast to the previous studies – on a highly aggregated subject level. Using all articles covered by WoS and published 1990, 1995, 2000, 2005, and 2010 we show the influence of the amount of linking on the variation of citation rates between disciplines. The data used in this analysis are from a bibliometrics database developed and maintained by the Max Planck Digital Library (MPDL, Munich) and derived from the Science Citation Index - Expanded (SCI-E), the Social Sciences Citation Index (SSCI), and the Arts and Humanities Citation Index (AHCI) produced by Thomson Reuters.

The graphs in figure 1 show the average number of citations per paper (bar), the average number of references per paper (gray line) and the average number of linked references per paper (black line), that is the fraction of references which are included in WoS in the form of database records. The results demonstrate that on average almost all disciplines show (more or less) similar average numbers of references in the appendices of these papers. The area of engineering and technology constitutes an exception, with a somewhat lower number (in all five years).

On the other hand, the average number of linked references varies widely: For example, in 2000 the number of linked references per paper was 1.66 in the humanities and 16.87 in the natural sciences.[1] The difference between the average total number of references and the average number of linked references is greatest in the case of the humanities, but also relatively large in the agricultural and the social sciences.

A comparison of the five graphs in Figure 1 shows that the pattern of dependency, at least over the selected period, hardly changed at all. However, the average number of citations of the seven disciplines analysed (grey bars) decreases continuously from 1990 to 2010 since the more recent papers had less time to accumulate citations. Simultaneously, the average number of references (triangles) and of linked references (squares) increased steadily.

Our results suggest – in agreement with previous studies – that the average citation rate is hardly influenced by the overall quantity of references, but far more by the extent to which the papers cited as references are included in WoS as database records. Since this extend is particularly low in the humanities, the average citation rate of humanity papers is correspondingly low. Consequently, one should consider less the citation culture or habits of the researchers in the different disciplines – in connection with the discipline-dependency of citations – and more the different coverage or comprehensiveness of the corresponding citation database.

---

[1] Inferential statistics (e.g. Analysis of Variance, ANOVA) is used in many empirical studies to analyse the statistical significance of group differences. In this study, we do not apply these statistics, because the huge publication numbers lead to statistically significant results despite only small citations differences.

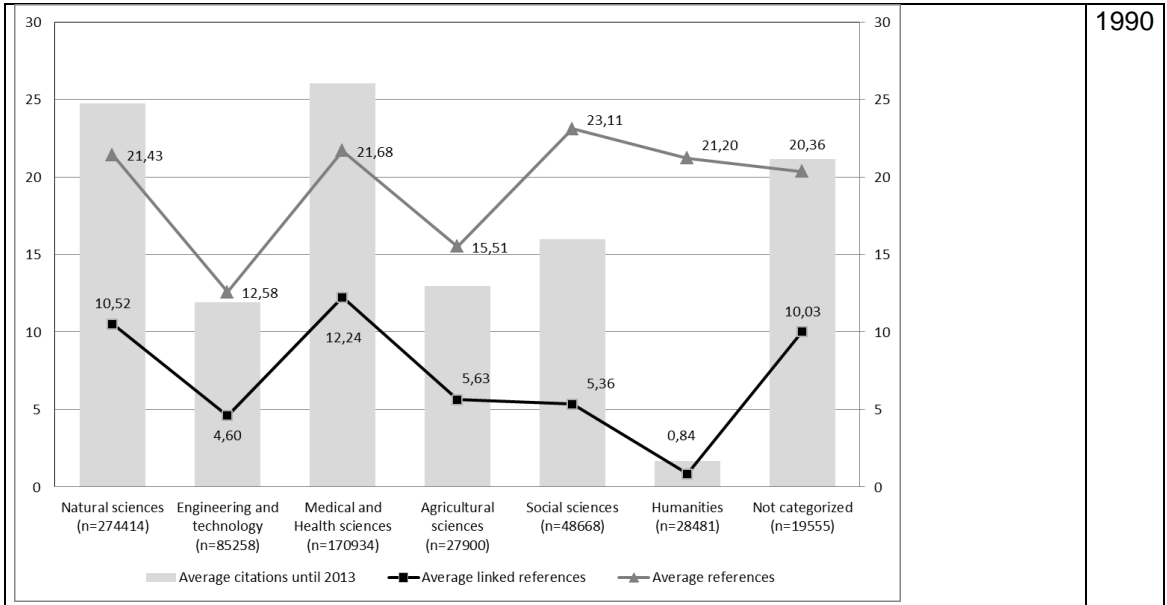

1990

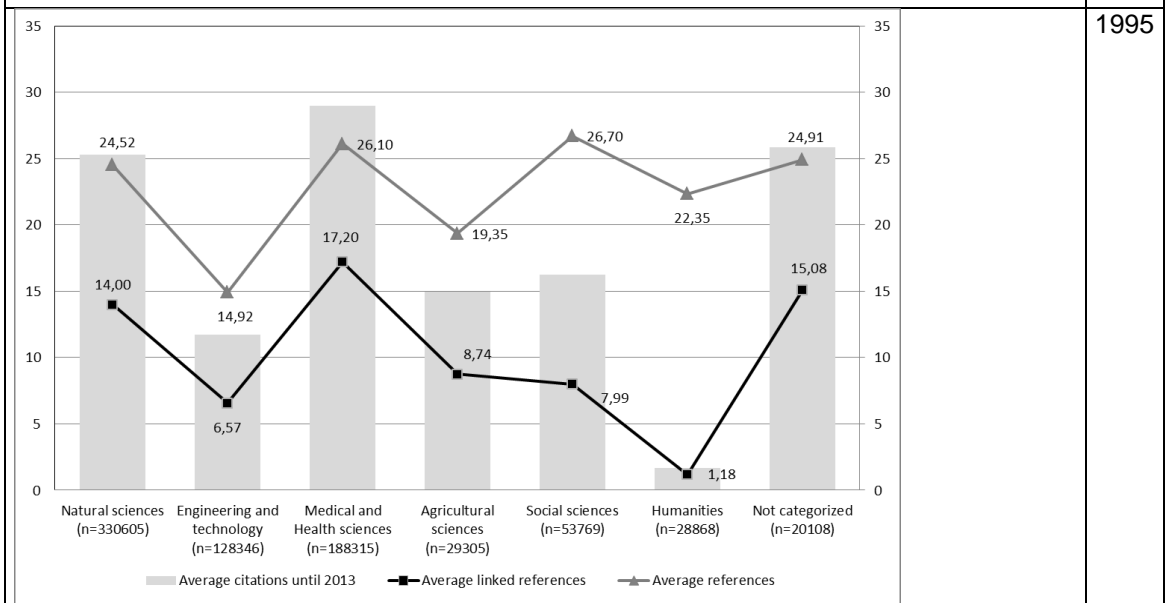

1995

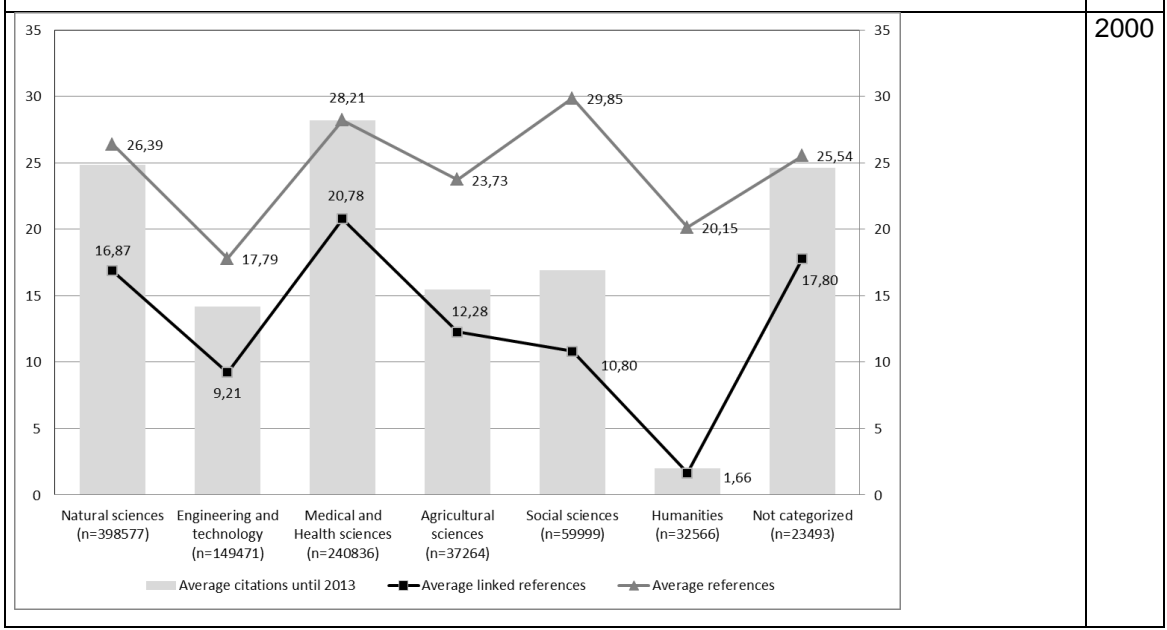

2000

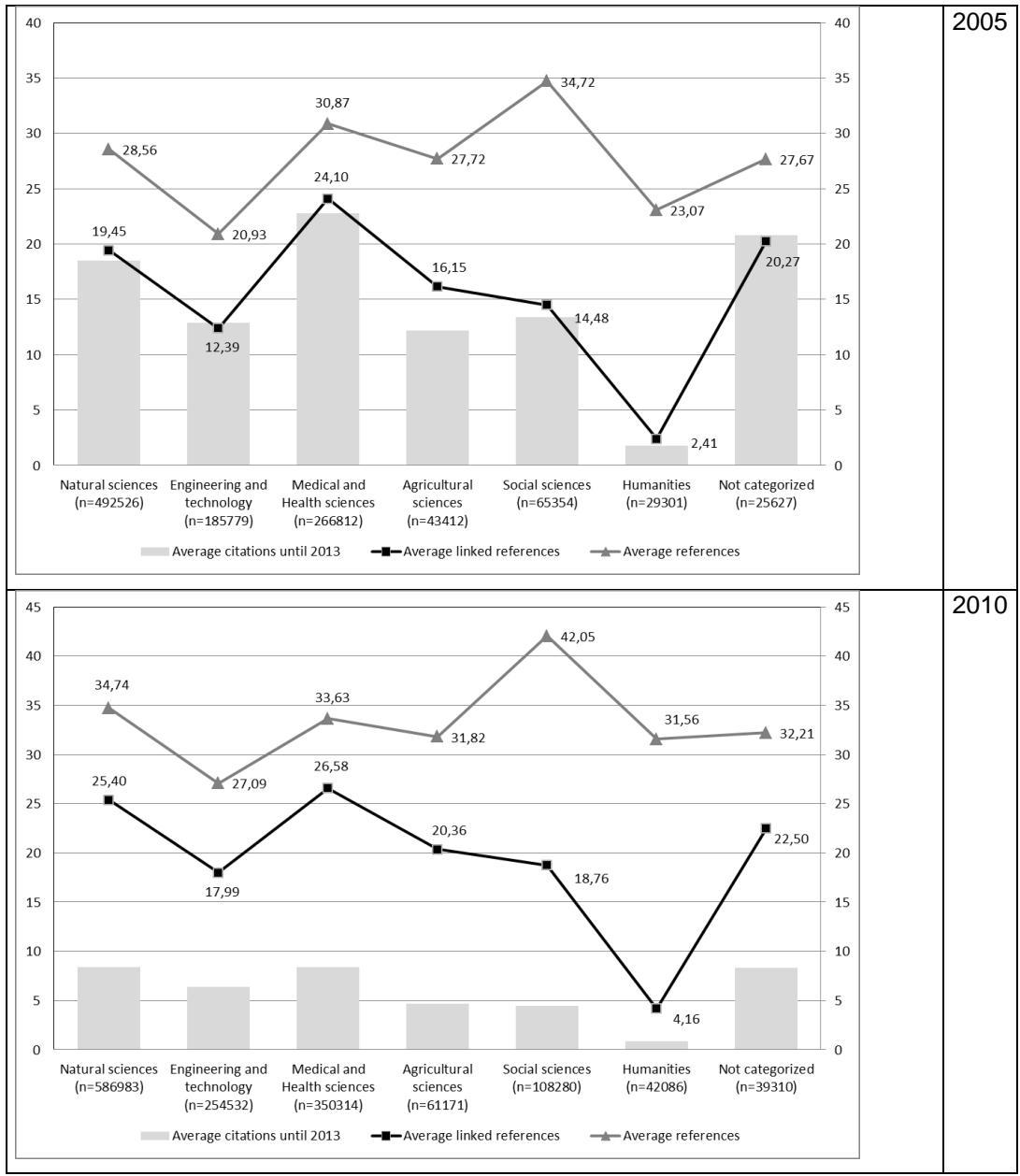

FIG 1: Average number of citations (grey bars), average number of cited references (triangles), and average number of linked cited references (squares) of articles published in 1990, 1995, 2000, 2005, and 2010. The articles have been categorized into disciplines by using the OECD Category scheme which corresponds to the Revised Field of Science and Technology (FOS) Classification of the Frascati Manual (Organisation for Economic Co-operation and Development).